\documentclass[%
 aip,
cp,  % Conference Proceedings
 amsmath,amssymb,%nobibnotes,
 reprint,%
]{revtex4-2}

\usepackage{graphicx}% Include figure files
\usepackage{dcolumn}% Align table columns on decimal point
\usepackage{bm}% bold math
\usepackage[utf8]{inputenc}
\usepackage[T1]{fontenc}
\usepackage{mathptmx} 
\usepackage{color}

\begin{document}

\title{Static analysis for coupled nonlinear Klein-Gordon equations with asymmetric parameter settings}% Force line breaks with \\

\author{Yasuhiro Takei} % Write as First name Surname
% \email[Corresponding author: ]{first.author@insitution.edu}
\affiliation{
  Mizuho Research $\&$ Technologies,Tokyo 101-8443,Japan % Force line breaks with \\ if necessary
}

\author{Yoritaka Iwata}%
 \email{Corresponding author: iwata$\_$phys@08.alumni.u-tokyo.ac.jp}
\affiliation{
  Kansai University,Osaka 564-8680,Japan % Force line breaks with \\ if necessary
}

%\author{Another's Name}
% \email{third.author@anotherinstitution.edu}
%\affiliation{%
%Second institution and/or address% Force line breaks with \\ if necessary
%}%
%\affiliation{You would list an author's second affiliation (if applicable) here.}

\date{\today} % It is always \today, today, but any date may be explicitly specified
              % Not printed for conference proceedings

\begin{abstract}
Klein-Gordon equations describe the dynamics of waves/particles in sub-atomic scales.
For a system of nonlinear Klein-Gordon equations, a systematic analysis of the time evolution for their spatially uniform solutions has been performed \cite{21takei}.
In the study, the parameters (mass, wave propagation speed, and the force parameters) are chosen to be symmetric between the two single equations.
Symmetric parameter settings are equivalent to assume the interacting two same particles.
In this paper, for a system of nonlinear Klein-Gordon equations with asymmetric parameter settings, the time evolution for their spatially uniform solutions are studied.
This is equivalent to assume the interacting two different particles.
As a result, based on the high precision numerical scheme \cite{22takei}, the existence of divergent and bounded solutions that depend on parameter settings is revealed.
The competition, coherence, and decoherence of different waves are shown to appear depending on the choice of asymmetrically-implemented parameter values.
\end{abstract}

\maketitle

%-----------------------------------------------------------------------------------------
%-----------------------------------------------------------------------------------------
%-----------------------------------------------------------------------------------------

\section{Introduction}
We consider one-dimensional wave equations with cubic nonlinearity.
Since the model equations correspond exactly to the $\phi^4$-theory in the quantum field theory (for a textbook, see \cite{65bjorken}), it is associated essentially with the Higgs mechanism.
For some nonlinear Klein-Gordon equations such as Sine-Gordon equations, the existence of breather solution has been known  \cite{93Denzler, 11Blank, 18Maier, 20Scheider}, and that solution are expected to behave as a kind of building unit of our universe. 
Indeed, the breather solution is a kind oscillation and behaves as time-periodic solution, so that it forms a kind of closed curve in the phase space.

In addition to the standard setting of breather solution, the periodic boundary condition is also imposed to the spatial direction $x$ in this study.
Therefore, a periodic solution for both time and space, which is likely to be called the space-time breather solution, is studied in nonlinear Klein-Gordon equation with cubic nonlinearity.
For the preceding result showing the existence of space-time periodic breather solution in a single Klein-Gordon equations, see \cite{1Takei}.

In this paper, fundamentals of the initial and boundary value problems for the coupled nonlinear Klein-Gordon equations with third-order nonlinearity are studied, keeping in mind investigating the existence of space-time periodic breather solutions in the future.
In particular, as a first step, the coupled nonlinear Klein-Gordon equation is studied in order to clarify the time evolution of the solution, which is uniform in the space direction, and several simplified ordinary differential equations are investigated to clarify the essence of the dynamical system.

The results confirm that by choosing symmetric or asymmetric parameters for the nonlinear Klein-Gordon equation system, there exist solutions that diverge or remain bounded depending on the initial values.
It should be noted here that the existence of bounded solutions with respect to the time evolution of spatially uniform solutions and its properties can provide important insights into the existence of spatially non-uniform breather solutions.

%-----------------------------------------------------------------------------------------
%-----------------------------------------------------------------------------------------
%-----------------------------------------------------------------------------------------

\section{Mathematical model}
Let $x \in [0,L]$ be a finite domain of space.
The evolution problem is studied for the positive time ($t \ge 0$). 
Let $\partial/\partial t$ and $\partial/\partial x$ be denoted by $\partial_t$ and $\partial_x$ respectively.
We begin with coupled nonlinear Klein-Gordon equations with the cubic nonlinearity, 
\[
\label{eq01}
\begin{array}{ll}
\ \partial_t^{2} u + \alpha_{1} \partial_x^{2} u - m_{1} u + k_{1} u^{3} + k^{\prime}_{1} u^{2} w = 0, \vspace{3mm} \\
\ \partial_t^{2} w + \alpha_{2} \partial_x^{2} w - m_{2} w + k_{2} w^{3} + k^{\prime}_{2} u w^{2} = 0, \vspace{3mm} \\
\ u(x, 0) = f(x),\ u(0, t) = u(L, t), \vspace{3mm} \\
\ \partial_t u(x, 0) = 0,\ \partial_t u(0, t) = \partial_t u(L, t), \vspace{3mm} \\
\ w(x, 0) = g(x),\ w(0, t) = w(L, t), \vspace{3mm} \\
\ \partial_t w(x, 0) = 0,\ \partial_t w(0, t) = \partial_t w(L, t),
\end{array}
\qquad {\rm (QT)}  
\]
where $\alpha_{1}, \alpha_{2}, m_{1}, m_{2}, k_{1}, k_{2}, k^{\prime}_{1}$ and $k^{\prime}_{2}$ are real constants. 
In addition to the third order self-interactions $u^3$ and $w^3$, mutual interactions $u^2 w$ and $uw^2$ of the same order are introduced.
The initial functions $f(x)$ and $g(x)$ are given as $L^2$-functions, and the periodic boundary condition is imposed for $x$-direction.
The breather solution has been shown to exist for Eq.~(\rm QT), if it only consists of the single equation (i.e., $k_1' =\alpha_2 = m_2 = k_2 = k_2' = 0$) \cite{1Takei}. 

In terms of clarifying fundamental units of coexisting states of breather solution, let us limit ourselves to the cases when $\alpha_{1} = \alpha_{2} = 0$.
The master equations are reduced to a system of ordinary differential equations
\[
\label{eq02}
\begin{array}{ll}
\ u_{tt} = m_{1} u - k_{1} u^{3} - k^{\prime}_{1} u^{2} w, \vspace{3mm} \\
\ w_{tt} = m_{2} w - k_{2} w^{3} - k^{\prime}_{2} u w^{2}.
\end{array}
\qquad {\rm (QT')}  
\]
For this ordinary differential equations, the initial values are given by real numbers; we take $u(x,0)=f(x)=u_{0}$, $w(x,0)=g(x)=w_{0}$, and $\partial_t u(x, 0) = \partial_t w(x, 0) = 0$.
In the following, we make systematics by choosing the values of $u_0$ and $w_0$.
Here $u_{tt}$ means the second order derivative of $u$ with respect to a variable $t$, and the same notation is true for $w_{tt}$.
%Note that Eq.~${\rm (QT^{\prime})}$ is essentially the same as the master equation of stationary problem
%\[
%\label{eq03}
%\begin{array}{ll}
%\ \alpha_1 u_{xx} - m_{1} u + k_{1} u^{3} + k^{\prime}_{1} u^{2} w = 0, \vspace{3mm} \\
%\ \alpha_2 w_{xx} - m_{2} w + k_{2} w^{3} + k^{\prime}_{2} u w^{2} = 0
%\end{array}
%\]
%with nonzero $\alpha_1$ and $\alpha_2$.

For a further simplification, the coefficients are fixed to be : $m_{1} = m_{2} = 1,\ k_{1} = k_{2} = 1,\ k^{\prime}_{1} = 1,\ k^{\prime}_{2} = \eta ( > 0)$.
In this paper the initial value problem ${\rm (QT^{\prime \prime})}$ is investigated.
Consequently, the wave propagation in the interacting two media is symmetric for the exchange between $u$ and $w$ ($\eta = 1$) and asymmetric ($0 < \eta \neq 1$).
This symmetric and asymmetric cases will provide a good starting point of analyzing complex interaction of coupled nonlinear Klein-Gordon equations. 
\[
\label{eq03}
\begin{array}{ll}
\ u_{tt} = u - u^{3} - u^{2} w, \vspace{3mm} \\
\ w_{tt} = w - w^{3} - \eta u w^{2}.
\end{array}
\qquad {\rm (QT^{\prime \prime})}  
\]

%-----------------------------------------------------------------------------------------
%-----------------------------------------------------------------------------------------
%-----------------------------------------------------------------------------------------

\section{Four special solutions\\ as essential pieces of the dynamical system}
%-----------------------------------------------------------------------------------------
\subsection{Cases with $u_{0}=0$ or $w_{0}=0$}
Much attention is paid to clarify the sensitivity to the initial condition.
For the time evolution of the solution of Eq.~${\rm (QT^{\prime \prime})}$, we see the different time evolution depending on the choice of initial value $(u_{0}, w_{0})$.
Let us begin with the cases with $w_0 = 0$.
The right-hand side of the second equation of Eq.~${\rm (QT^{\prime \prime})}$ is represented by
\begin{equation}
\label{eq04}
w - w^{3} - \eta u w^{2} = w( 1 - w^{2} - \eta u w),
\end{equation}
so that $w=0$ is the fixed point of the second equation.
That is, $w(t)=0,\ t >0$ is a stationary solution, and it is realized if we take $w_{0}=0$.
On the other hand, by substituting $w(t)=0$, the right-hand side of the first equation of Eq.~${\rm (QT^{\prime \prime})}$ can be represented by
\begin{equation}
 u - u^{3}.
\end{equation}
In this situation Eq.~${\rm (QT^{\prime \prime})}$ is reduced to
\[
\begin{array}{ll}
\ u_{tt} = u - u^{3}, \quad  %\vspace{3mm} \\
\ w = 0.
\end{array}
\qquad {\rm (QT^{\prime \prime}_{1})}
\label{eq05}
\]
The solutions of (${\rm QT^{\prime \prime}_{1}}$) with the initial condition $u(0) = u_{0}$ and $u_{t}(0) = 0$ are necessarily located on the horizontal axis $(u, 0)$ in the $u-w$ plane, and all we have to do is to solve the single equation $u_{tt} = u - u^{3}$.
For the behavior of the solution to this single equation, see \cite{1Takei}.

%Equation ${\rm (QT^{\prime \prime})}$ is symmetric for the exchange between $u$ and $w$.
By the same idea as above, in case with $u_{0}=0$, Eq.~${\rm (QT^{\prime \prime})}$ is also reduced to the single equation.
\[
\begin{array}{ll}
\ u = 0, \quad %\vspace{3mm} \\
\ w_{tt} = w - w^{3},
\end{array}
\qquad {\rm (QT^{\prime \prime}_{2})}  
\label{eq06}
\]
The solutions of (${\rm QT^{\prime \prime}_{2}}$) with the initial condition $w(0) = w_{0}$ and $w_{t}(0) = 0$ are necessarily located on the vertical axis $(0, w)$ in the $u-w$ plane.
The solutions of  (${\rm QT^{\prime \prime}_{1}}$) and (${\rm QT^{\prime \prime}_{2}}$) play a role of basic units in the phase space analysis.

%-----------------------------------------------------------------------------------------
\subsection{Cases with $w_{0}= c u_{0}$}
Let us move on to the cases with initial value satisfying $u_{0}= c w_{0}$, where $c$ is a constant.
Substituting $w = c u$ into the second equation of the master equation of Eq.~${\rm (QT^{\prime \prime})}$, we obtain
\[
\begin{array}{ll}
\ u_{tt} = u - (c^{2} + \eta c ) u^{3}. \quad %\vspace{3mm} \\ 
\end{array}
\]
Here, considering that this equation is identical to the first equation of Eq.~${\rm (QT^{\prime \prime})}$, we have $c^2 + \eta c = 1 + c$. 
Therefore, $c$ can be expressed as $c_{\pm} = \frac{1 - \eta}{2} \pm \sqrt{(\frac{\eta-1}{2})^{2} + 1}$.

When $c_{+}$ is chosen, the solution is necessarily represented by $(u, c_{+} u)$, and the master equation of Eq.~${\rm (QT^{\prime \prime})}$ is expressed as an essentially-single equation.
%The solution is necessarily represented by $(u, c u)$, and the master equation of Eq.~${\rm (QT^{\prime \prime})}$ is expressed as an essentially-single equation.
\[
\begin{array}{ll}
\ u_{tt} = u - (1 + c_{+}) u^{3}, \quad %\vspace{3mm} \\ 
\ w = c_{+} u.
\end{array}
\qquad {\rm (QT^{\prime \prime}_{3})}  
\label{eq07}
\]
Since $c_{+} > 0,\ (\eta > 0)$, also $1+c_{+}$ is  positive.
Therefore, the solutions of (${\rm QT^{\prime \prime}_{3}}$) with the initial condition $u(0) = u_{0}$ and $u_{t}(0) = 0$ are necessarily located on the line $u = c_{+} w$ in the $u-w$ plane.
Similarity of the master equation to ${\rm (QT^{\prime \prime}_{1})}$ and ${\rm (QT^{\prime \prime}_{2})}$ are noticed in this case.

In the same manner, let us consider the case with $u_{0} = c_{-} w_{0}$. 
The solution is represented by $(u, c_{-} u)$ in the $u-w$ plane, and the master equation of Eq.~${\rm (QT^{\prime \prime})}$ is also expressed as an essentially-single equation.
\[
\begin{array}{ll}
\ u_{tt} = u - (1 + c_{-}) u^{3}, \quad %\vspace{3mm} \\ 
\ w = c_{-} u.
\end{array}
\qquad {\rm (QT^{\prime \prime}_{4})}  
\label{eq08}
\]
The solutions of (${\rm QT^{\prime \prime}_{4}}$) with the initial condition $u(0) = u_{0}$ and $u_{t}(0) = 0$ are necessarily located on the line $u= c_{-} w$ in the $u-w$ plane.
It should be noted here that, unlike ${\rm (QT^{\prime \prime}_{1})}$, ${\rm (QT^{\prime \prime}_{2})}$ and ${\rm (QT^{\prime \prime}_{3})}$ cases, depending on the value of $\eta$, $u$ diverges in infinite time (cf. grow-up of solution).

Specifically, when $0 < \eta \leq 1$, since $0 \leq \frac{1 - \eta}{2} < \frac{1}{2}$, then
\[
\begin{array}{ll}
1+c_{-} \ = \ \frac{1-\eta}{2} + 1 - \sqrt{(\frac{1-\eta}{2})^{2} + 1} %\vspace{3mm} \\
\ \geq \ (\frac{1-\eta}{2})^{2} + 1 - \sqrt{(\frac{1-\eta}{2})^{2} + 1} %\vspace{3mm} \\
\ \geq \ 0. %\quad %\vspace{3mm} \\ 
\end{array}
\]
Note that $1+c_{-}=0$ is the case when $\eta = 1$.
On the other hand, when $\eta > 1$,
\[
\begin{array}{ll}
1+c_{-} \ &= \ \frac{1-\eta}{2} + 1 - \sqrt{(\frac{1-\eta}{2})^{2} + 1} %\vspace{3mm} \\
\ < \ 1 - \sqrt{(\frac{1-\eta}{2})^{2} + 1} %\vspace{3mm} \\
\ < \ 0. %\quad %\vspace{3mm} \\ 
\end{array}
\]
Therefore, $1+c_{-}$ is negative or zero when $\eta \ge 1$, so the solutions of (${\rm QT^{\prime \prime}_{4}}$) diverges in infinite time.
Since $1+c_{-}$ is positive when $0 < \eta < 1$, the solutions of (${\rm QT^{\prime \prime}_{4}}$) with the initial condition $u(0) = u_{0}$ and $u_{t}(0) = 0$ are necessarily located on the line $u = c_{-} w$ in the $u-w$ plane.

%-----------------------------------------------------------------------------------------
\subsection{Cross section of dynamical system}
\begin{figure}[tb] 
\begin{center}
\includegraphics[width=80mm,bb=60 0 300 196]{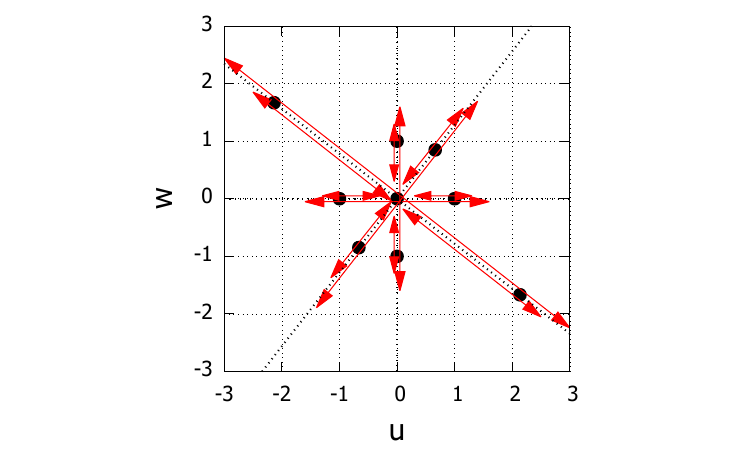}
  \caption{\label{fig01a}Dynamical system of ${\rm (QT')}$ ($\eta$ = 0.5) is shown in $u-w$ plane. 
Four model cases are shown by arrows (shown as arrows), where red thick arrows mean oscillatory motions.
The black points are the fixed points.
Solution orbits of ${\rm (QT'_{1})}$, ${\rm (QT'_{2})}$, ${\rm (QT'_{3})}$, and ${\rm (QT'_{4})}$ show the oscillatory motion around the fixed points.
}
\end{center}
\end{figure}

\begin{figure}[tb] 
\begin{center}
\includegraphics[width=80mm,bb=60 0 300 196]{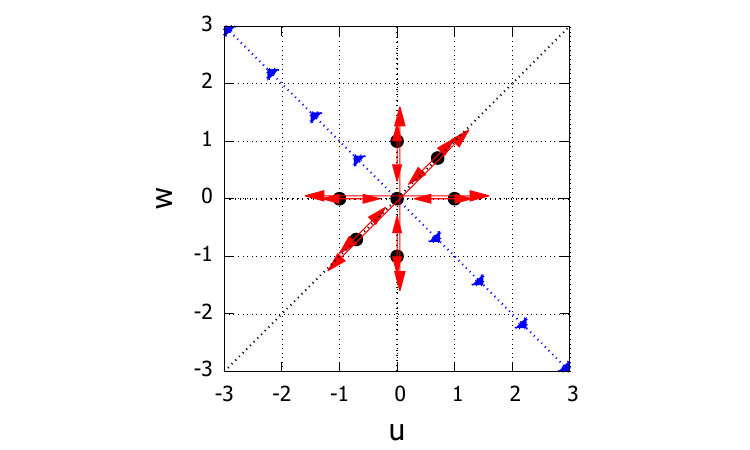} 
\includegraphics[width=80mm,bb=60 0 300 196]{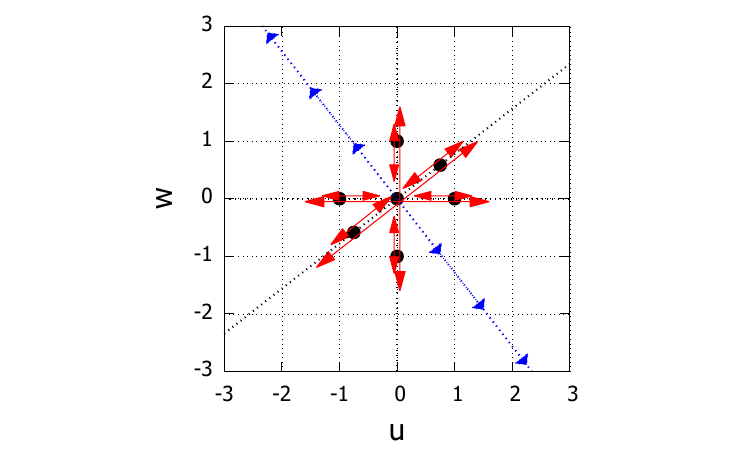}
  \caption{\label{fig01b} Dynamical system of  ${\rm (QT')}$ ($\eta$ = 1.0) is shown in $u-w$ plane (the left panel). Dynamical system of  ${\rm (QT')}$ ($\eta$ = 1.5) is shown in $u-w$ plane (the right panel).
Four model cases are shown by arrows (shown as arrows), where red thick arrows mean oscillatory motions, and blue dashed arrows show the  divergent motions.
The black points are the fixed points.
Solution orbits of ${\rm (QT'_{1})}$, ${\rm (QT'_{2})}$, and ${\rm (QT'_{3})}$ show the oscillatory motion around the fixed points.
Solution orbits of ${\rm (QT'_{4})}$ diverge monotonically to $(+\infty,-\infty)$ or $(-\infty,+\infty)$.
}
\end{center}
\end{figure}

All the four model cases (${\rm QT^{\prime \prime}_{1}}$) to (${\rm QT^{\prime \prime}_{4}}$) are shown by four lines in Fig.~\ref{fig01a}($\eta = 0.5$) and Fig.~\ref{fig01b}($\eta = 1, 1.5$).
Solutions of (${\rm QT^{\prime \prime}}$) with general initial data evolve between the four lines in Fig.~\ref{fig01a} and Fig.~\ref{fig01b}.
Note that in order to complete the dynamical system representation, we have to add two additional variables $\partial_t u$ and $\partial_t w$ in Fig.~\ref{fig01a} and Fig.~\ref{fig01b}, so that two-dimensional cross section of the full four-dimensional dynamical system is shown in Fig.~\ref{fig01a} and Fig.~\ref{fig01b}.
%Here, when $\eta = 1$, the dynamical system, which consists of the phase space and the solution orbits, becomes symmetric with respect to the line $u=-w$.
%This fact arises from the symmetric setting of coefficients $\eta$.
The purpose of this paper is to draw detail structures inside Fig.~\ref{fig01a} and Fig.~\ref{fig01b}.

%-----------------------------------------------------------------------------------------
%-----------------------------------------------------------------------------------------
%-----------------------------------------------------------------------------------------
\section{Solutions with general initial data}
%-----------------------------------------------------------------------------------------
\subsection{Divergent and bounded solution}
The time evolution of the solution to Eq.~${\rm (QT^{\prime \prime})}$ has a wide variety of cases other than the four fundamental cases.
Based on the four model cases, solution orbits starting from general initial values are studied.
Much attention is paid to find the initial condition to hold bounded solutions. 
In the present paper, the initial values of first derivative is always given by $u_t(0)=w_t(0)=0$.

\begin{figure}[tb]
\begin{center}
\includegraphics[width=65mm,bb=75 0 285 216]{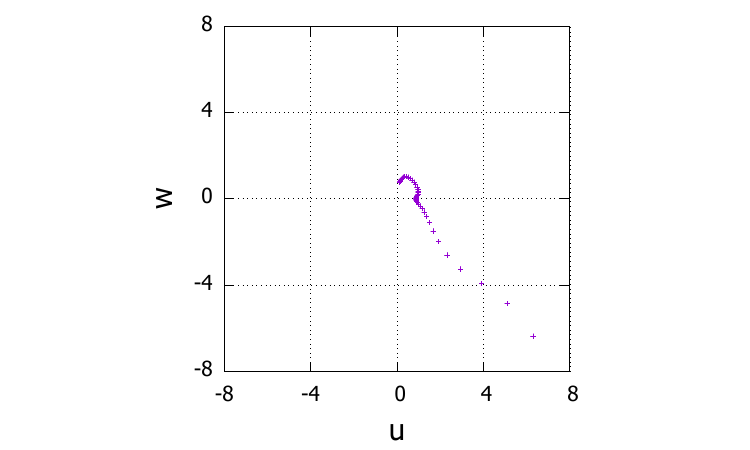}
\includegraphics[width=65mm,bb=75 0 285 216]{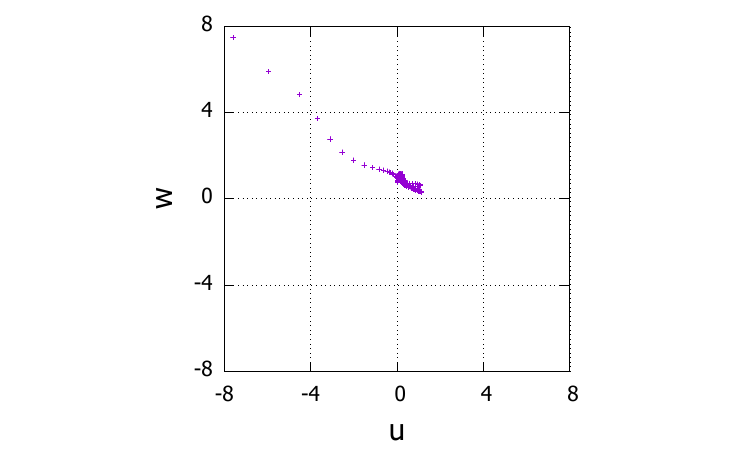}
  \caption{($\eta = 1$ case) Trajectory of the divergent solutions $(u, w),\ (0 \le t \le 1024)$. 
Left and right panels correspond to the cases with the initial value $(u_{0}, w_{0})=(0.125, 0.75)$ and that with the initial value $(u_{0}, w_{0})=(0.025, 0.75)$, respectively.}
\label{fig02}
\end{center}
\begin{center}
\includegraphics[width=65mm,bb=75 0 285 216]{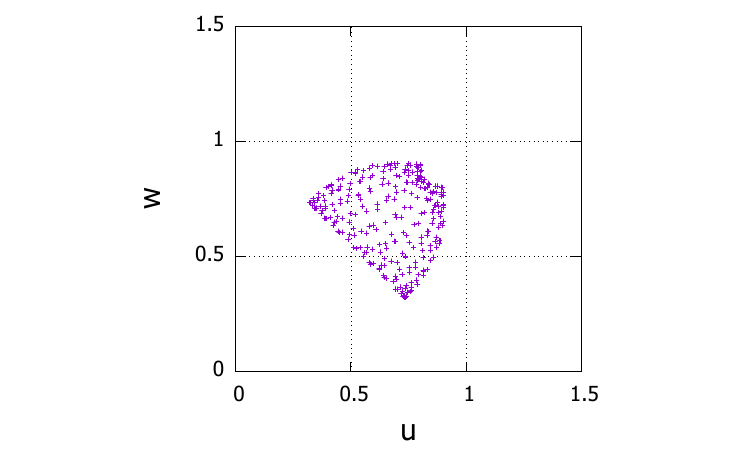}
\includegraphics[width=65mm,bb=75 0 285 216]{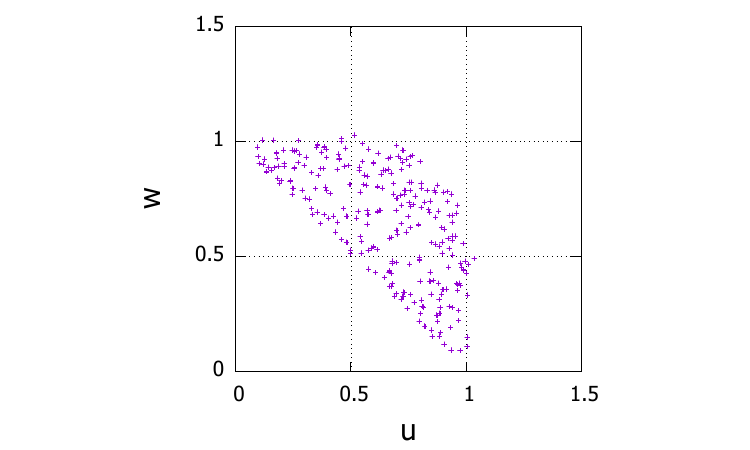}
  \caption{($\eta = 1$ case) Trajectory of bounded solutions $(u, w) \ (0 \le t \le 1024)$. 
Left and the right panels correspond to the cases with the initial value $(u_{0}, w_{0})=(0.8, 0.9)$ and that with the initial value $(u_{0}, w_{0})=(0.1, 0.9)$, respectively.}
\label{fig03}
\end{center}
\end{figure}

The solution $u(t)$ is called divergent if there exists a certain positive real number $t_M < \infty$ such that   
\[ |u(t_M)| > M \]
for an arbitrary real number $M > 0$.
On the other hand, the solution $u(t)$ is called bounded if there exists a certain positive real number $M < \infty$ such that   
\[ \max_{t \in [0, \infty)} | u(t) | \le M. \]
Depending on the initial data, divergent/bounded solutions appear.
In this paper, these solutions are examined with a sufficiently long time interval $t \in [0, 1024]$.

For example, as shown in the left panel of  Fig.~\ref{fig02} \ ($\eta = 1$ case), the solution diverges when the initial value is chosen as $(u_{0}, w_{0})=(0.125, 0.75)$.
Similarly, as shown in the right panel of  Fig.~\ref{fig02} \ ($\eta = 1$ case), the solution diverges when the initial value is chosen as $(u_{0}, w_{0})=(0.025, 0.75)$.
Here it is clear that, depending on the initial settings, some solutions possibly diverge.
By comparing the left and right panels of Fig.~\ref{fig02} \ ($\eta = 1$ case), small difference results in a different kind of divergence.
The coupled system inherently holds the sensitivity to the initial data; one goes to the upper left side, and the other goes to the lower right side.

Depending on the choice of initial data, some solutions of Eq.~${\rm (QT^{\prime \prime})}$ remain in a finite domain. 
Those solutions are called the bounded solutions in this paper.
%%%
For example, as shown in the left panel of Fig.~\ref{fig03} \ ($\eta = 1$ case), the solution is bounded when the initial value is chosen as $(u_{0}, w_{0})=(0.8, 0.9)$.
Similarly, as shown in the right panel of Fig.~\ref{fig03} \ ($\eta = 1$ case), the solution is also bounded when the initial value is chosen as $(u_{0}, w_{0})=(0.1, 0.9)$.
Although the examination has been done for a finite duration time, no divergent modes appear in these cases, and modes like time-periodic oscillation appear instead.
It is a supportive evidence for the existence of bounded solutions.

%-----------------------------------------------------------------------------------------
\subsection{Distribution of divergent and bounded solutions on the phase space}
Here, for each of the three cases $\eta = 0.5, 1, 1.5$, we systematically calculate the numerical solution orbits started from general initial values, and find the distribution of initial values on the phase space where divergent and bounded solutions occur .

Specifically, the initial value $(u_{0}, w_{0})$ is set as the cross point of the mesh that divides $N_{1}$ in the $u$ axes and $N_{2}$ in the $w$ axes, respectively, for a region $-8 \le u \le 8,\ -8 \le w \le 8$ in the $u-v$ plane.
For that initial value, the time evolution of the solution of Eq.~${\rm (QT^{\prime \prime})}$ up to time $t=T$ is calculated and determined whether the solution is divergent or bounded by the relationship between $\max_{t \in [0,T]} |u(T)|, \max_{t \in [0,T]} |w(T)|$ for some value $M$.
Below are the calculation results for each of the three cases $\eta = 0.5,\ 1,\ 1.5$.

%-----------------------------------------------------------------------------------------
\subsubsection{The case of $\eta=0.5$}
Here we consider the solution of Eq.~${\rm (QT^{\prime \prime})}$ when $\eta = 0.5$.
Before looking at the numerical results, first, in order to grasp the structure of the phase space ($u-v$ plane), let $f(u, w)=u-u^{3}-u^{2}w,\ g(u, w)=w-w^{3}-\eta u w^{2}$ for the right side of Eq.~${\rm (QT^{\prime \prime})}$, and consider the point $(u, w)$ satisfying $f(u, w)=0,\ g(u, w)=0$.

\begin{figure}[tb] 
\begin{center}
\includegraphics[width=80mm,bb=60 0 300 196]{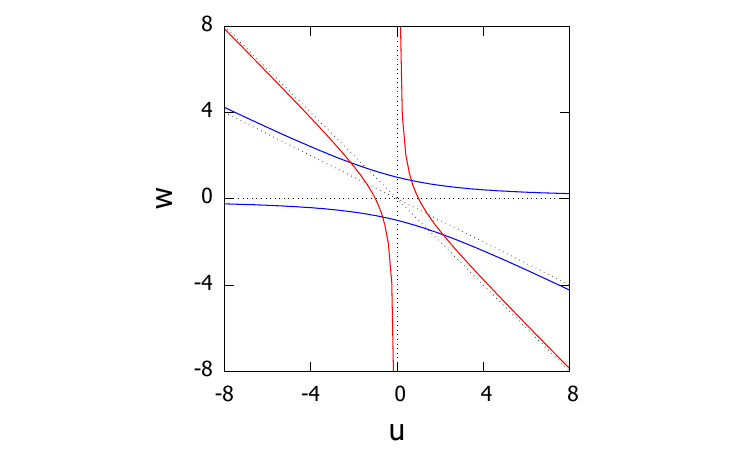} 
\includegraphics[width=80mm,bb=60 0 300 196]{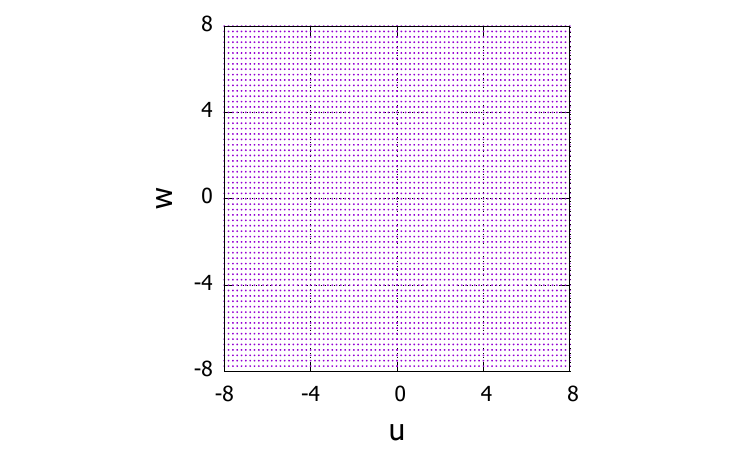} 
  \caption{\label{fig04a}
(Left panel) The red curve shows $(u,w)$ satisfying $u_{tt}=f(u,w)=0$. In the respective regions of $u>0$ and $u<0$, $u_{tt}=f(u,w)>0$ at the point $(u,w)$ to the left of the red curve, and $u_{tt}=f(u,w)<0$ at the point $(u,w)$ to the right.
The blue curve shows $(u,w)$ satisfying $w_{tt}=g(u,w)=0$. In the respective regions of $w>0$ and $w<0$, $w_{tt}=g(u,w)>0$ at the point $(u,w)$ below the blue curve, and $w_{tt}=g(u,w)<0$ at the point $(u,w)$ above.\\
(Right panel) Distribution of initial values where bounded and divergent solutions occur in the region $-8 \le u \le 8,\ -8 \le w \le 8$ on the $u-v$ plane. The purple dots indicate the distribution of initial values where bounded solutions occur, and the green dots indicate the distribution of initial values where divergent solutions occur.
}
\end{center}
\end{figure}

To satisfy $f(u,w)=u-u^{3}-u^{2}w=u (1-u^{2}-uw)=0$, $u=0$ or $(u, w)$ must satisfy the following equation when $u \neq 0$.
\begin{equation}
w=\frac{(1-u)(1+u)}{u}.
\label{eq09}
\end{equation}
Next, to satisfy $g(u,w)=w-w^{3}-\eta w^{2}u=w (1-w^{2}-\eta wu)=0$, $w=0$ or $(u, w)$ must satisfy the following equation when $w \neq 0$.
\begin{equation}
w= -\frac{\eta u}{2} \pm \sqrt{(\frac{\eta u}{2})^{2} + 1}.
\label{eq10}
\end{equation}
The equations (\ref{eq09}) and (\ref{eq10}) are drawn on the $u-v$ plane with red and blue curves, respectively, as shown in Fig.\ref{fig04a} left.
Here, the four intersections of the blue and red curves as well as $(u, w)=(0,0)$ are points that simultaneously satisfy $f(u, w)=0,\ g(u, w)=0$, and the solution with those initial values is an fixed point. (Note that this paper assumes $u_{t}(0)=0, w_{t}(0)=0$ as initial values.)

For points other than the above, in the region $u<0$, $u_{tt}=f(u,w) > 0$ for points to the left of the red curve, and $u_{tt}=f(u,w) < 0$ for points to the right of the red curve.
Similarly, in the region $u > 0$, $u_{tt}=f(u,w) > 0$ at points to the left of the red curve, and $u_{tt}=f(u,w) < 0$ at points to the right of the red curve. At $u=0$, $u_{tt}=f(u,w)=0$.
In the region $w<0$, $w_{tt}=g(u,w)>0$ at points below the blue curve, and $w_{tt}=g(u,w)<0$ at points above the blue curve.
Similarly, in the region $w>0$, $w_{tt}=g(u,w)>0$ at points below the blue curve, and $w_{tt}=g(u,w)<0$ at points above the blue curve. At $w=0$, $w_{tt}=g(u,w)=0$.
From the above, when $\eta = 0.5$, it is inferred that the solution with initial values at points on the $u-v$ plane transitions around the four intersections of the blue and red curves and $(u, w)=(0,0)$.

While taking the above into consideration, the distribution of initial values where bounded and divergent solutions occurred is shown as Fig. \ref{fig04a} right.
Here, the purple dots in the figure indicate the distribution of initial values where bounded solutions occur, and the green dots indicate the distribution of initial values where divergent solutions occur.
Note that $N_{1} = 80,\ N_{2} = 64$ as the numerical setting, and the region $-8 \le u \le 8,\ -8 \le w \le 8$ on the $u-v$ plane is divided into $80 \times 64$.
Numerical calculations were systematically performed up to time $T=1024$ using $(u_{0}, w_{0})$ as initial values for a total of $5120$ points (see \cite{22takei, 21IwataTakei} for numerical methods, etc.).
As can be seen from this result, when $\eta = 0.5$, the solution for every divided points on the domain $-8 \le u \le 8,\ -8 \le w \le 8$, with that point as the initial value, is a bounded solution.

\begin{figure}[tb] 
\begin{center}
\includegraphics[width=80mm,bb=60 0 300 196]{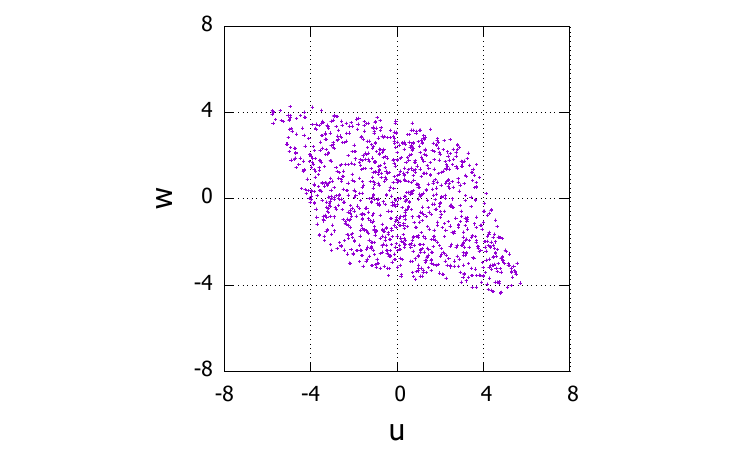} 
  \caption{\label{fig04a4}
Trajectory of bounded solution for initial value $(u_{0}, w_{0})=(1.2,\ 2.75)$.
The solution trajectory is seen to be spread evenly in the diamond-shaped region including the four intersections of the blue and red curves and $(u, w)=(0,0)$ in the left diagram of Fig. \ref{fig04a}.
}
\end{center}
\end{figure}

Next, we present the characteristics of a typical solution trajectory for a bounded solution.
A solution with initial values at points on the domain $-8 \le u \le 8,\ -8 \le w \le 8$ is a bounded solution. As an example of a typical solution trajectory, the trajectory of the solution for initial values $(u_{0}, w_{0})=(1.2,\ 2.75)$ is shown in Fig. \ref{fig04a4}.
As can be seen in this figure, the trajectory of the solution is evenly spread over the area on the rhombus, including the four intersections of the blue and red curves and $(u, w)=(0,0)$ (the fixed point) in the left figure of Fig. \ref{fig04a}.

\begin{figure}[tb] 
\begin{center}
\includegraphics[width=80mm,bb=60 0 300 196]{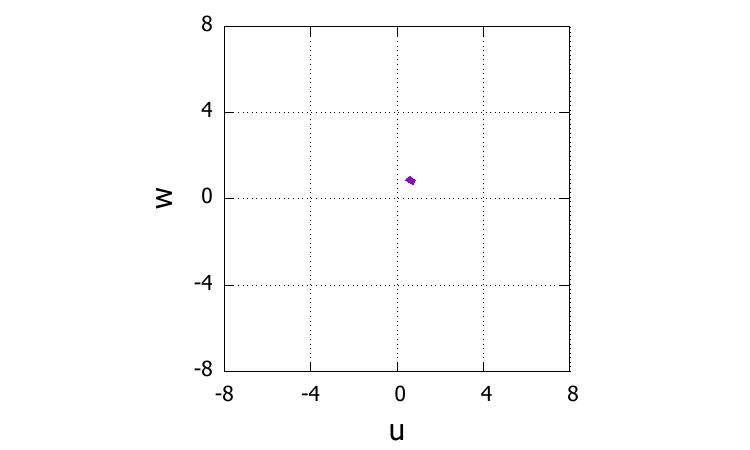} 
\includegraphics[width=80mm,bb=60 0 300 196]{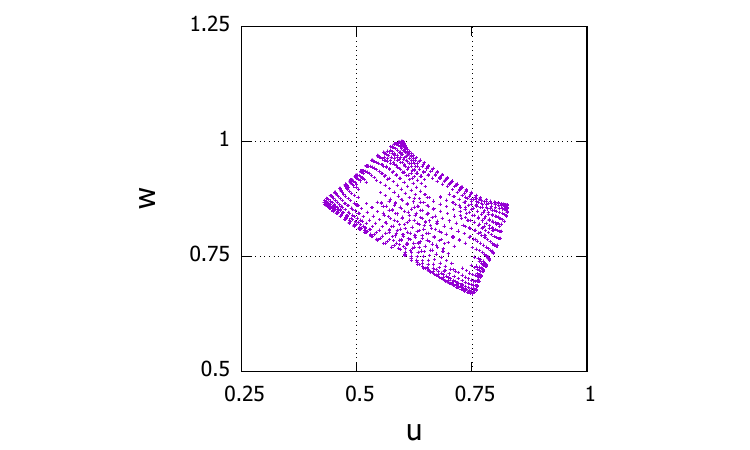} 
  \caption{\label{fig04a1}
Trajectory of bounded solution with initial value of $(u_{0}, w_{0})=(0.8,\ 6.25)$.
The left figure shows the solution trajectory within the entire region $-8 \le u \le 8,\ -8 \le w \le 8$.
The figure on the right zooms in on the solution trajectory to get a detailed view of the solution trajectory. The braid-like trajectory can be seen within the specific range of $0.3< u <0.8,\ 0.6< w <1.1$.
}
\end{center}
\end{figure}

On the other hand, in addition to the typical bounded solution trajectories described above, there are also cases in which the trajectories are geometrically distinctive.
For example, the trajectory of the bounded solution is shown in Fig. \ref{fig04a1} when the initial value is $(u_{0}, w_{0})=(0.6,\ 1.0)$.
The left of Fig. \ref{fig04a1} shows the solution trajectory within the entire region $-8 \le u \le 8,\ -8 \le w \le 8$, and in particular, we can see how the solution trajectory is restricted to a specific range where $u,\ w >0$.
Furthermore, in order to understand the solution trajectory in detail, the solution trajectory is zoomed in on the right side of Fig. \ref{fig04a1}.
This shows that the solution has periodic motion within a specific range of limited $0.3<u<0.8,\ 0.6<w<1.1$.

\begin{figure}[tb] 
\begin{center}
\includegraphics[width=80mm,bb=60 0 300 196]{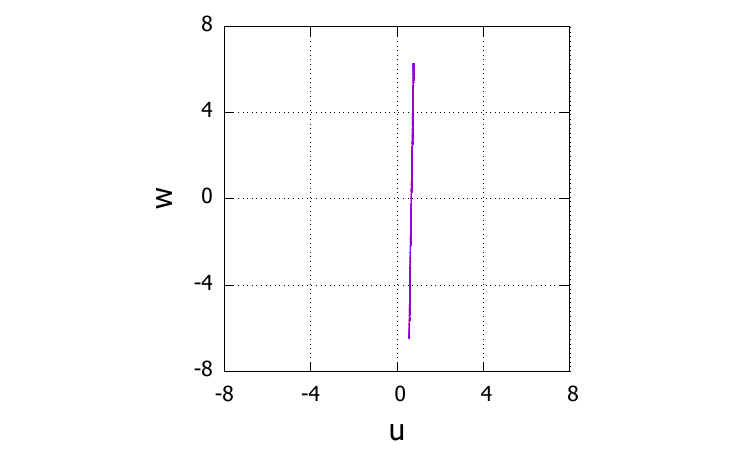} 
\includegraphics[width=80mm,bb=60 0 300 196]{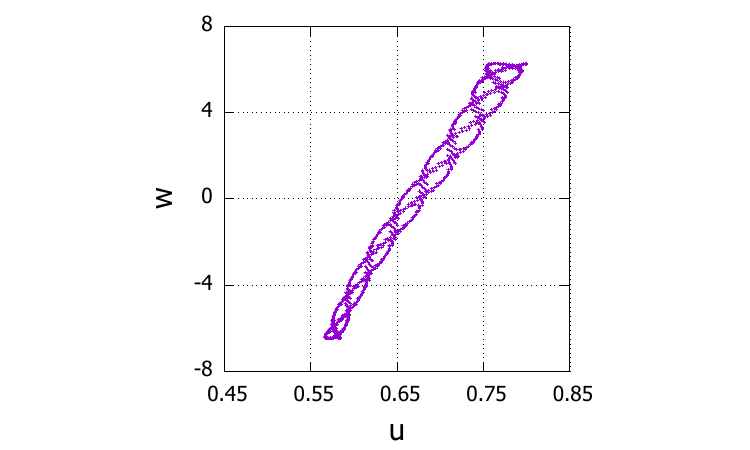} 
  \caption{\label{fig04a3}
Trajectory of bounded solutions with initial values of $(u_{0}, w_{0})=(0.6,\ 1.0)$.
The left figure shows the solution trajectory within the entire region $-8 \le u \le 8,\ -8 \le w \le 8$.
The figure on the right zooms in on the solution trajectory to get a detailed view of the solution trajectory. We can see periodic movements within a specific range where $0.55<u<0.85,\ -8<w<8$.
}
\end{center}
\end{figure}

The trajectory of the bounded solution is shown in Fig. \ref{fig04a3} when the initial value is $(u_{0}, w_{0})=(0.8,\ 6.25)$.
Similar to the above, the left figure of Fig. \ref{fig04a3} shows the trajectory of the solution within the entire region $-8 \le u \le 8,\ -8 \le w \le 8$, and it can be seen that the trajectory of the solution is limited to a specific range where $u >0$.
In particular, the change of the solution in the $w$-axis direction is larger than that in the $u$-axis direction, and the solution trajectory can be understood as a round trip on a straight line almost parallel to the $w$-axis in the $u-v$-plane.
Furthermore, in order to grasp the solution trajectory in detail, a zoomed-in view of the solution trajectory is shown on the right in Fig. \ref{fig04a3}.
This shows that within a specific range of $0.55< u <0.85,\ -8< w <0$, the trajectory is like a braid.

Although the graphical notation is omitted here, it has been confirmed that there exist solutions with periodic motion similar to that of Fig. \ref{fig04a1} within the specific range of $-0.8<u<-0.3,\ -1.1<w<-0.6$, as a trajectory similar to that of the characteristic bounded solution described above.
Also, within the specific range of $-0.85< u <-0.55,\ -8< w <8$, we have confirmed a solution that draws a braid-like trajectory similar to the right figure in Fig. \ref{fig04a3}.
Furthermore, we have also confirmed a solution that draws a trajectory similar to the case where $u$ and $w$ are interchanged for the trajectory in the right figure of Fig.\ref{fig04a3}.

%\begin{figure}[tb] 
%\begin{center}
%\includegraphics[width=80mm,bb=60 0 300 196]{eta050_u060_f0030_075_large.pdf} 
%\includegraphics[width=80mm,bb=60 0 300 196]{eta050_u060_f0030_075_small.pdf} 
%  \caption{\label{fig04a2} comment
%}
%\end{center}
%\end{figure}

%-----------------------------------------------------------------------------------------
\subsubsection{The case of $\eta=1.0$ and $\eta=1.5$}
\begin{figure}[tb] 
\begin{center}
\includegraphics[width=80mm,bb=60 0 300 196]{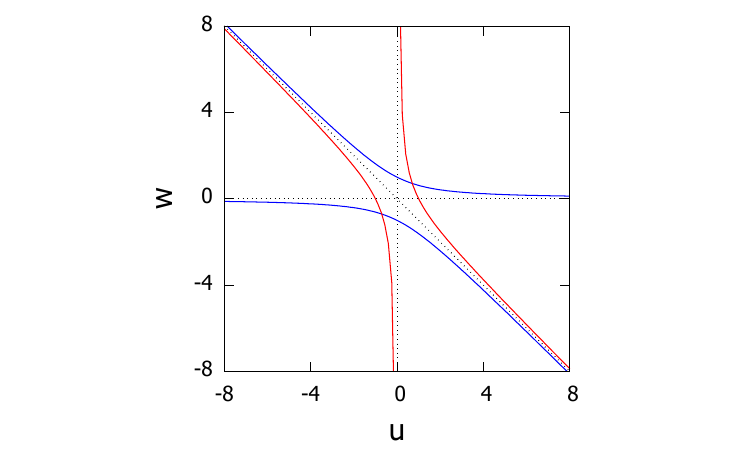} 
\includegraphics[width=80mm,bb=60 0 300 196]{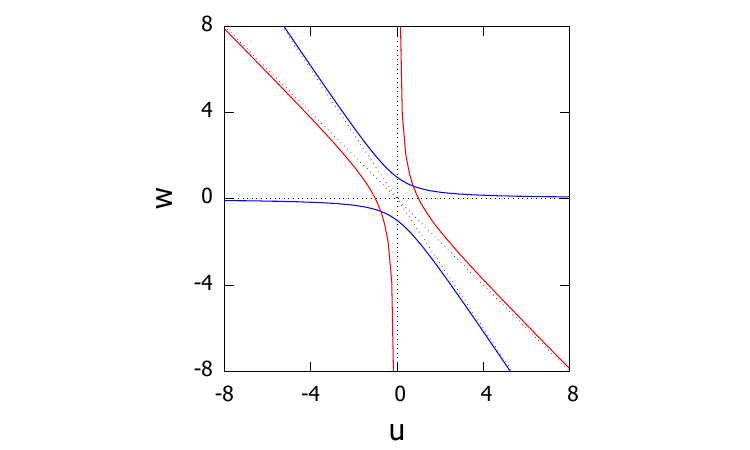} 
  \caption{\label{fig05a}
The left figure shows the curve for $\eta=1.0$ and the right figure shows the curve for $\eta=1.5$.
In each figure, the red curve shows $(u,w)$ satisfying $u_{tt}=f(u,w)=0$. In the respective regions of $u>0$ and $u<0$, $u_{tt}=f(u,w)>0$ at the point $(u,w)$ to the left of the red curve, and $u_{tt}=f(u,w)<0$ at the point $(u,w)$ to the right.
In each figure, the blue curve shows $(u,w)$ satisfying $w_{tt}=g(u,w)=0$. In the respective regions of $w>0$ and $w<0$, $w_{tt}=g(u,w)>0$ at the point $(u,w)$ below the blue curve, and $w_{tt}=g(u,w)<0$ at the point $(u,w)$ above.
}
\end{center}
\end{figure}

Then consider the solution of Eq.~${\rm (QT^{\prime \prime})}$ when $\eta = 1.0,\ 1.5$.
As in the previous section, before looking at the numerical results, in order to grasp the structure of the phase space ($u-v$-plane), let $f(u, w)=u-u^{3}-u^{2}w,\ g(u, w)=w-w^{3}-\eta u w^{2 }$, and consider the point $(u, w)$ satisfying $f(u, w)=0,\ g(u, w)=0$.
The equations (\ref{eq09}) and (\ref{eq10}) with $\eta = 1.0,\ 1.5$ are drawn on the $u-v$ plane with red and blue curves respectively, Fig.\ref{fig05a}.

Here, the intersection points of the blue and red curves as well as $(u, w)=(0,0)$ are points that simultaneously satisfy $f(u, w)=0,\ g(u, w)=0$, and the solution with those initial values is a fixed point. (Note that this paper assumes $u_{t}(0)=0, w_{t}(0)=0$ as initial values.)
However, compared to the case of $0<\eta<1$, the blue and red curves intersect at only two points.
The difference from the $0<\eta<1$ case is that the blue and red curves do not intersect in the $u>0,\ w<0$ and $u<0,\ w>0$ regions.

For points other than the above, in the region $u<0$, $u_{tt}=f(u,w) > 0$ for points to the left of the red curve, and $u_{tt}=f(u,w) < 0$ for points to the right of the red curve.
Similarly, in the region $u > 0$, $u_{tt}=f(u,w) > 0$ at points to the left of the red curve, and $u_{tt}=f(u,w) < 0$ at points to the right of the red curve. At $u=0$, $u_{tt}=f(u,w)=0$.
In the region $w<0$, $w_{tt}=g(u,w)>0$ at points below the blue curve, and $w_{tt}=g(u,w)<0$ at points above the blue curve.
Similarly, in the region $w>0$, $w_{tt}=g(u,w)>0$ at points below the blue curve, and $w_{tt}=g(u,w)<0$ at points above the blue curve. At $w=0$, $w_{tt}=g(u,w)=0$.
From these facts, if $\eta = 1.0,\ 1.5$, in the region $u>0,\ w<0$, the point $(u, w)$ to the left of the red curve and above the blue curve is inferred to evolve in time to $(+\infty, -\infty)$.
Similarly, in the region $u<0,\ w>0$, the point $(u, w)$ to the right of the red curve and below the blue curve is presumed to evolve in time to $(-\infty, +\infty)$.
This shows that the structure of the phase space ($u-v$ plane) differs between $0< \eta < 1$ and $\eta \geq 1$.

\begin{figure}[tb] 
\begin{center}
\includegraphics[width=80mm,bb=60 0 300 196]{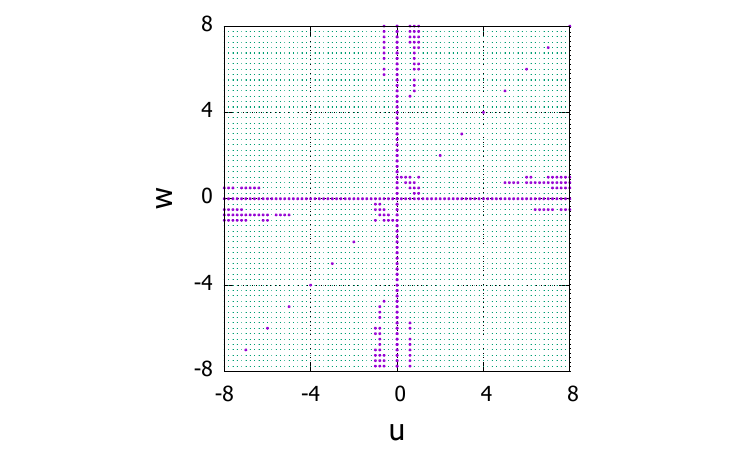} 
\includegraphics[width=80mm,bb=60 0 300 196]{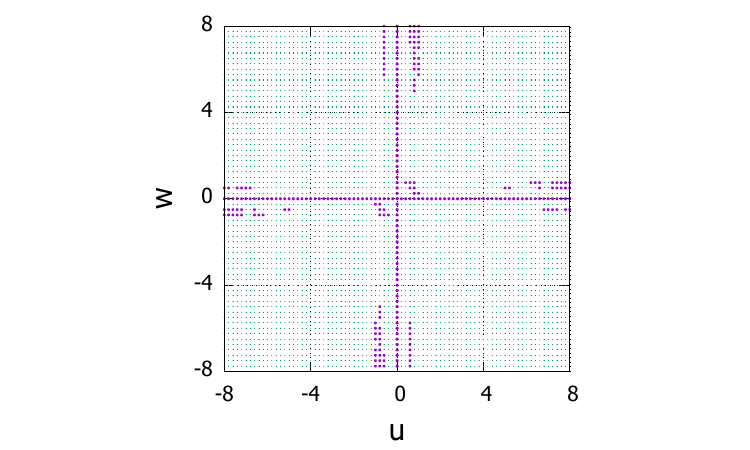} 
  \caption{\label{fig05b}
The left figure shows the case $\eta =1$ and the right figure shows the case $\eta =1.5$.
The distribution of initial values where bounded and divergent solutions occur in the region $-8 \le u \le 8,\ -8 \le w \le 8$ on the $u-v$ plane. The purple dots indicate the distribution of initial values where bounded solutions occur, and the green dots indicate the distribution of initial values where divergent solutions occur.
}
\end{center}
\end{figure}

While taking the above into consideration, the distribution of initial values where bounded and divergent solutions occurred for each case of $\eta = 1.0,\ 1.5$ is shown in Fig. \ref{fig05b}.
Here, the purple dots in the figure indicate the distribution of initial values where bounded solutions occur, and the green dots indicate the distribution of initial values where divergent solutions occur.
Note that $N_{1} = 80,\ N_{2} = 64$ as the numerical setting, and the region $-8 \le u \le 8,\ -8 \le w \le 8$ on the $u-v$ plane is divided into $80 \times 64$.
Numerical calculations were systematically performed up to time $T=1024$ using $(u_{0}, w_{0})$ as initial values for a total of $5120$ points.
As can be seen from the results, when $\eta = 1.0,\ 1.5$, the solution is divergent at many points set on the region $-8 \le u \le 8,\ -8 \le w \le 8$, and only solutions with initial values at some region points are bounded.

\begin{figure}[tb] 
\begin{center}
\includegraphics[width=80mm,bb=60 0 300 196]{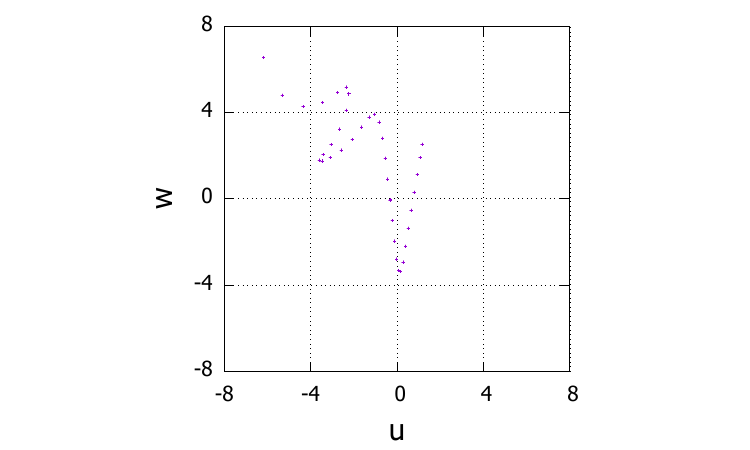} 
  \caption{\label{fig06}
Trajectory of divergent solution for initial value $(u_{0}, w_{0})=(1.2,\ 2.75)$.
The point $(u, w)$ can be seen to evolve (diverge) in time to $(-\infty, +\infty)$.
}
\end{center}
\end{figure}

In the following, we will show the characteristics of typical solution trajectories for divergent and bounded solutions, taking the case of $\eta=1.5$ as representative.
Solutions with many points on the domain $-8 \le u \le 8,\ -8 \le w \le 8$ (Fig. \ref{fig05b}) as initial values are divergent solutions, and as an example of a typical solution trajectory for the initial value $(u_{0}, w_{0})=(1.2,\ 2.75)$, Fig.\ref{fig06}.
As you can see in this figure, the point $(u, w)$ is seen to evolve (diverge) in time to $(-\infty, +\infty)$.

\begin{figure}[tb] 
\begin{center}
\includegraphics[width=80mm,bb=60 0 300 196]{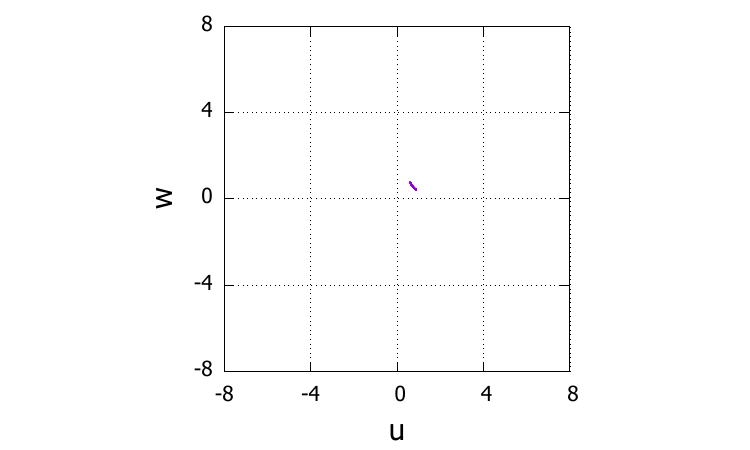} 
\includegraphics[width=80mm,bb=60 0 300 196]{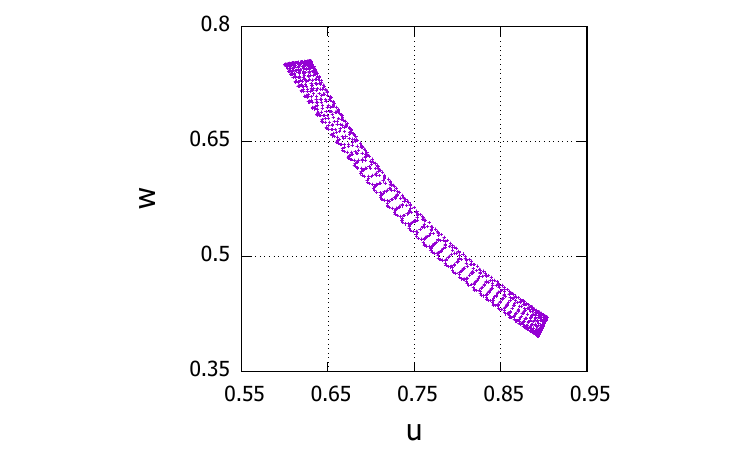} 
  \caption{\label{fig06a1}
Trajectory of bounded solution with initial value of $(u_{0}, w_{0})=(0.6,\ 0.75)$.
The left figure shows the solution trajectory within the entire region $-8 \le u \le 8,\ -8 \le w \le 8$.
The figure on the right zooms in on the solution trajectory to get a detailed view of the solution trajectory. 
We can see periodic movements within a specific range where $0.6<u<0.95,\ 0.35<w<0.8$.
}
\end{center}
\end{figure}

On the other hand, in addition to the above divergent solutions, there are also solutions that are bounded solutions and draw geometrically distinctive trajectories.
For example, the trajectory of the bounded solution is shown in Fig. \ref{fig06a1} when the initial value is $(u_{0}, w_{0})=(0.6,\ 0.75)$.
The left of Fig. \ref{fig06a1} shows the solution trajectory within the entire region $-8 \le u \le 8,\ -8 \le w \le 8$, and in particular, we can see how the solution trajectory is restricted to a specific range where $u,\ w >0$.
Furthermore, in order to grasp the solution trajectory in detail, the solution trajectory is zoomed in on the right side of Fig. \ref{fig06a1}.
This shows that the solution has a periodic movement within a specific range of limited $0.6<u<0.95,\ 0.35<w<0.8$.

\begin{figure}[tb] 
\begin{center}
\includegraphics[width=80mm,bb=60 0 300 196]{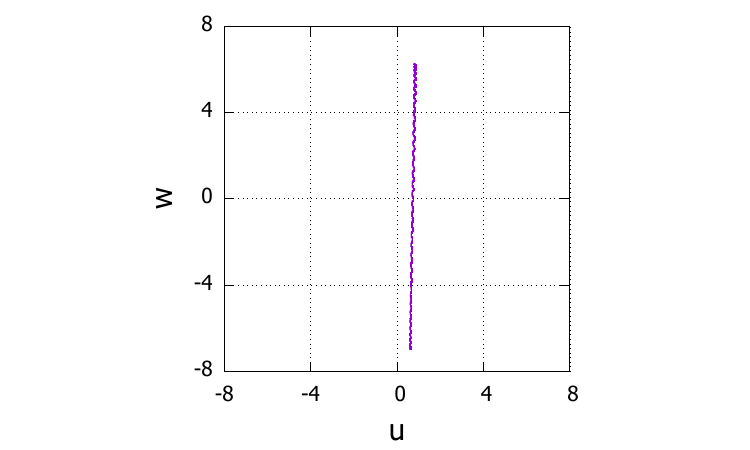} 
\includegraphics[width=80mm,bb=60 0 300 196]{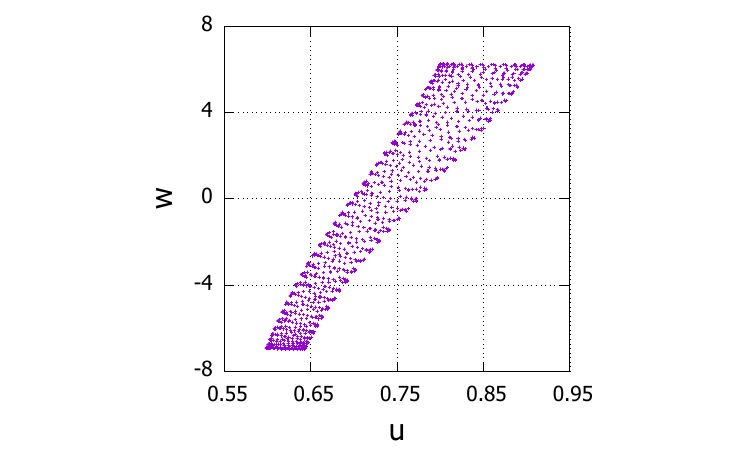} 
  \caption{\label{fig06a3}
Trajectory of bounded solution with initial value of $(u_{0}, w_{0})=(0.8,\ 6.25)$.
The left figure shows the solution trajectory within the entire region $-8 \le u \le 8,\ -8 \le w \le 8$.
The figure on the right zooms in on the solution trajectory to get a detailed view of the solution trajectory. 
We can see periodic movements within a specific range where $0.55< u <0.95,\ -8< w <8$.
}
\end{center}
\end{figure}

The trajectory of the bounded solution is shown in Fig. \ref{fig06a3} when the initial value is $(u_{0}, w_{0})=(0.8,\ 6.25)$.
Similar to the above, the left of Fig. \ref{fig06a3} shows the trajectory of the solution within the entire region $-8 \le u \le 8,\ -8 \le w \le 8$, and it can be seen that the trajectory  of the solution is limited to a specific range where $u >0$. 
In particular, the change of the solution in the $w$-axis direction is larger than that in the $u$-axis direction, and the solution trajectory can be understood as a round trip on a straight line almost parallel to the $w$-axis in the $u-v$-plane.
Furthermore, in order to understand the solution trajectory in detail, the solution trajectory is zoomed in on the right side of Fig. \ref{fig06a3}.
It can be seen that within the specific range of $0.55< u <0.95,\ -8< w <8$, the periodic movement can be seen.

Although the graphical representation is omitted here, it has been confirmed that there exist solutions with periodic motion similar to that of Fig. \ref{fig06a1} within the specific range of $-0.95<u<-0.6,\ -0.8<w<-0.35$.
In addition, within the specific range of $-0.95< u <-0.55,\ -8< w <8$, we have also confirmed a solution that draws a periodic trajectory similar to the right of Fig. \ref{fig06a3}.
Furthermore, we have also confirmed a solution that draws a trajectory similar to the case where $u$ and $w$ are interchanged for the trajectory in the right of Fig.\ref{fig06a3}.

%-----------------------------------------------------------------------------------------
\section{Discussion}
Klein-Gordon equations describe the dynamics of waves/particles in sub-atomic scales.
In this paper, for a system of nonlinear Klein-Gordon equations with asymmetric parameter settings, the time evolution for their spatially uniform solutions were studied.
This is equivalent to assume the interacting two different particles.

By systematically analyzing the trajectories of the solutions with respect to the initial values, we found that, in particular, whether a divergent  solution occurs depends on whether the parameter $\eta>0$ is smaller or larger than 1.
Among them, we also found that there are several bounded solutions with geometrically characteristic trajectories.
In particular, the characteristics of the bounded solutions can be classified into the following categories.

\begin{quote}
\begin{itemize}
\item Solutions where both $u$ and $w$ continue to be positive or both $u$ and $w$ continue to be negative.
\item Solution where $u$ continues to take positive (or negative) values while $w$ cyclically takes positive and negative values.
\item Solution where $w$ continues to take positive (or negative) values while $u$ cyclically takes positive and negative values.
\end{itemize}
\end{quote}
The characteristic bounded solutions organized above can also be considered to correspond to certain attractors on the phase space($u-v$ plane).
Therefore, if we can consider spatially non-uniform solutions that connect the above multiple attractors after setting appropriate boundary conditions, etc., we can expect that they can be space-time breather solutions as discussed in \cite{1Takei}.

%-----------------------------------------------------------------------------------------
\section{Conclusion}
In this paper, fundamentals of the initial and boundary value problems for the coupled nonlinear Klein-Gordon equations with third-order nonlinearity were studied, keeping in mind investigating the existence of space-time periodic breather solutions in the future.
In particular, as a first step, the coupled nonlinear Klein-Gordon equations were studied in order to clarify the time evolution of the solution, which is uniform in the space direction, and several simplified ordinary differential equations were investigated to clarify the essence of the dynamical system.

The results confirm that by choosing symmetric or asymmetric parameters for the coupled nonlinear Klein-Gordon equations, there exist solutions that diverge or remain bounded depending on the initial values.
It should be noted here that the existence of bounded solutions with respect to the time evolution of spatially uniform solutions and its properties can provide important insights into the existence of spatially non-uniform breather solutions.

In particular, for solutions $u, w$ corresponding to the behavior of two different types of interacting particles, it was found that one solution continues to hold positive (or negative) values, while the other has positive and negative values that are interchangeable.
It was also found that there exist solutions where both continue to hold positive (or negative) values.

These results show the behavior of spatially uniform solutions, but when we consider the existence of spatially non-uniform breather solutions based on \cite{1Takei}, they suggest the existence of solutions that are breather solutions for one particle and oscillatory solutions for the other particle.  It also implies the existence of two particles that are simultaneously a breather solution.

In the future, by analyzing the spatially non-uniform solutions of the coupled nonlinear Klein-Gordon equations with third-order nonlinearity with reference to the results of this paper, it is expected that the existence of breather solutions for two types of interacting particles will be shown.

%-----------------------------------------------------------------------------------------
%-----------------------------------------------------------------------------------------
%\begin{acknowledgments}
%We wish to acknowledge the support of the author community in using
%REV\TeX{}, offering suggestions and encouragement, testing new versions,
%\dots.
%\end{acknowledgments}

\nocite{*}
\bibliography{aipsamp}% Produces the bibliography via BibTeX.

\end{document}